\documentclass[twocolumn]{aastex631}

\hypersetup{linkcolor=red,citecolor=blue,filecolor=cyan,urlcolor=blue}

\usepackage{graphicx,color}
\usepackage{amssymb}
\usepackage{amsmath}
\usepackage{url}
\usepackage{natbib}
\usepackage{txfonts}
\usepackage{multirow}
\usepackage{array}
\usepackage{rotating}
\usepackage{sidecap}
\usepackage{hyperref}
\usepackage{epstopdf}
\usepackage{footnote}
\usepackage{tabularx}
\usepackage{booktabs}
\usepackage{longtable}
\usepackage{tabu}

\newcommand{\FeX}{\ion{Fe}{10}}
\newcommand{\FeIX}{\ion{Fe}{9}}

\newcommand{\sdo}{\textit{SDO}}

\newcommand{\hri}{HRI$_{EUV}$}

\shortauthors{Panesar et al.}

\accepted{: ApJ Letters}

\shorttitle{Magnetic Origin of Campfires Observed by SO and SDO}
\shortauthors{Panesar et al.}

\begin{document}
	\title{The Magnetic Origin of Solar Campfires}

\correspondingauthor{Navdeep K. Panesar}
\email{panesar@lmsal.com}

\author[0000-0001-7620-362X]{Navdeep K. Panesar}
\affil{Lockheed Martin Solar and Astrophysics Laboratory, 3251 Hanover Street, Bldg. 252, Palo Alto, CA 94304, USA}
\affil{Bay Area Environmental Research Institute, NASA Research Park, Moffett Field, CA 94035, USA}

	\author[0000-0001-7817-2978]{Sanjiv K. Tiwari}
\affil{Lockheed Martin Solar and Astrophysics Laboratory, 3251 Hanover Street, Bldg. 252, Palo Alto, CA 94304, USA}
\affil{Bay Area Environmental Research Institute, NASA Research Park, Moffett Field, CA 94035, USA}

	\author[0000-0003-4052-9462]{David Berghmans}
	\affil{Solar-Terrestrial Centre of Excellence – SIDC, Royal Observatory of Belgium, Ringlaan -3- Av. Circulaire, 1180 Brussels, Belgium}

\author[0000-0003-2110-9753]{Mark C. M. Cheung}
\affil{Lockheed Martin Solar and Astrophysics Laboratory, 3251 Hanover Street, Bldg. 252, Palo Alto, CA 94304, USA}

\author[0000-0001-9027-9954]{Daniel M{\"u}ller}
\affil{European Space Agency, ESTEC, P.O. Box 299, 2200 AG Noordwijk, The Netherlands} 

\author[0000-0003-0972-7022]{Frederic Auchere}
\affil{Universit\'{e} Paris-Saclay, CNRS, Institut d’Astrophysique Spatiale, 91405, Orsay, France}

\author[0000-0002-2542-9810]{Andrei Zhukov}
\affil{Solar-Terrestrial Centre of Excellence – SIDC, Royal Observatory of Belgium, Ringlaan -3- Av. Circulaire, 1180 Brussels, Belgium}
\affil{Skobeltsyn Institute of Nuclear Physics, Moscow State University, 119992 Moscow, Russia}

\begin{abstract}
 Solar campfires  are  fine-scale heating events, recently observed by Extreme Ultraviolet Imager (EUI), onboard Solar Orbiter. Here we use  EUI 174\AA\ images, together with EUV images from \sdo/AIA, and line-of-sight magnetograms from \sdo/HMI to investigate the magnetic origin of 52 randomly selected campfires in the quiet solar corona. We find that (i) the campfires are rooted at the edges of  photospheric magnetic network lanes; (ii) most of the campfires reside above the neutral line between majority-polarity magnetic flux patch and a merging minority-polarity flux patch, with a flux cancelation rate of  $\sim$10$^{18}$Mx hr$^{-1}$; (iii) some of the campfires occur repeatedly from the same neutral line; (iv) in the large majority of instances, campfires are preceded by a cool-plasma structure, analogous to minifilaments in coronal jets; and (v) although many campfires have `complex' structure, most campfires resemble  small-scale jets, dots, or loops. Thus, `campfire' is a general term that includes different types of small-scale solar dynamic features. They contain sufficient magnetic energy ($\sim$10$^{26}$-10$^{27}$ erg) to heat the solar atmosphere locally to 0.5--2.5MK. Their lifetimes range from about a minute to over an hour, with most of the campfires having a lifetime of $<$10 minutes. The average lengths and widths of the campfires are 5400$\pm$2500km and 1600$\pm$640km, respectively. Our observations suggest that (a) the presence of magnetic flux ropes may be ubiquitous in the solar atmosphere and not limited to coronal jets and larger-scale eruptions that make CMEs, and (b) magnetic flux cancelation is the fundamental process for the formation and triggering of most campfires.

\end{abstract}


\section{Introduction} \label{sec:intro}

The Solar Orbiter  mission was launched on 10 February 2020 \citep{muller2020}. The payload consists of six remote sensing and four in-situ instruments. One of its instruments, the Extreme Ultraviolet Imager (EUI; \citealt{rochus2020}) has two High Resolution Imagers (HRIs): \hri\ and HRI$_{Lya}$. Both the HRIs  observed the quiet solar corona, in the 174 \AA\ and 1216 \AA\ passbands, during the first  perihelion pass of Solar Orbiter.  The \hri\ captured small-scale brightenings, known as \textit{campfires}  \citep{berghmans2021}, on the solar disk center in 174 \AA\ images. 

Solar campfires are small-scale, short-lived, coronal brightenings, and can appear as loop-like, dot-like or complex structures. They are mostly rooted  at the chromospheric  network boundaries and their height lies between 1000 and 5000 km above the photosphere  \citep{berghmans2021,zhukov2021}. The formation mechanisms of campfires and their connection to the photospheric magnetic field are unknown.  

Campfires might be magnetic reconnection events in the quiet Sun corona, at small-scales, in low-lying magnetic structures. Therefore, these could intrinsically be similar to sub-flares (nanoflares, microflares) and larger-scale solar eruptions \citep{Parker88,priest00,asc04}. Consistently, \cite{chen21} proposed, using an MHD model, that campfires are mostly caused by component magnetic reconnection, heating the quiet Sun corona to 1 MK or more.

In this Letter, we examine the magnetic field evolution at the  base of campfires and investigate what triggers these heating events. We also examine the physical properties of campfires. For this purpose, we combine the \hri\  images with images of   \textit{Solar Dynamics Observatory} (\sdo)/Atmospheric Imaging Assembly (AIA; \citealt{lem12}), and investigate the photospheric magnetic field evolution of ondisk campfires using  co-aligned line of sight magnetograms from \sdo/Helioseismic and Magnetic Imager (HMI; \citealt{scherrer12}). By studying a total of 52 campfires observed at 44 different locations, we find that (i) 77\% (40 out of 52) of campfires  appear at sites of magnetic flux cancelation between  the majority-polarity flux patch and a merging minority-polarity  flux patch, and (ii) 79\% (41 out of 52)   of campfires are  accompanied by structures of cool plasma.

\begin{figure*}[ht]
	\centering
	\includegraphics[width=0.85\linewidth]{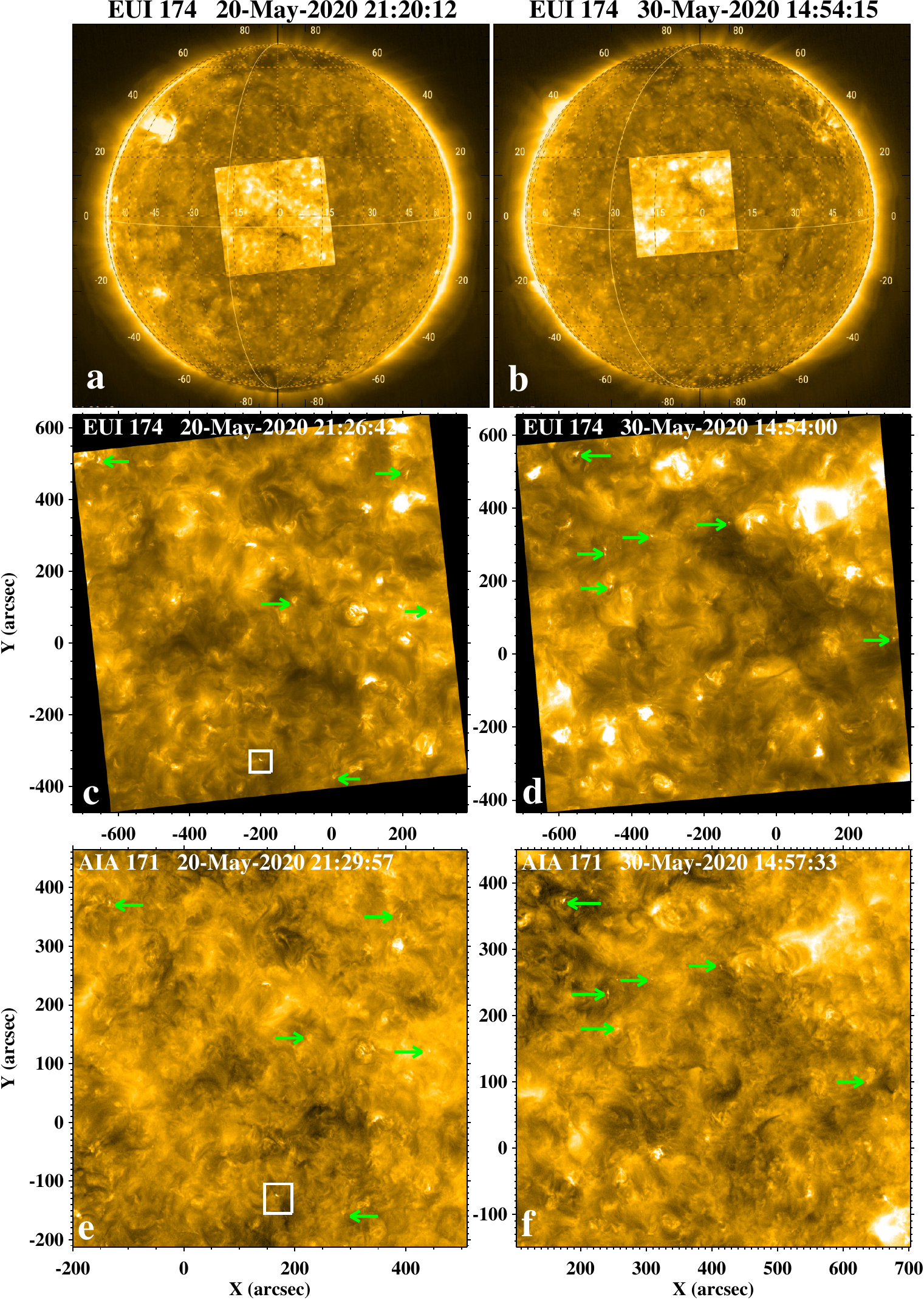} 
	\caption{Examples of campfires observed on 20-May-2020 and 30-May-2020. Panels (a) and (b) show the EUI/Full Sun Imager (FSI) 174 \AA\ images overlaid with the High Resolution Imager (HRI) field of view. The intersections of the two solid yellow circles in (a) and (b) indicate the SDO subsolar points. Panels (c) and (d) show the \hri\ images of the same region that is overlaid in panels (a) and (b), respectively. Panels (e) and (f) display AIA 171 \AA\ images of the same field of view shown in panels (c) and (d). The green arrows  (in c and d) point to some of the campfires that we have studied. The green arrows in the AIA images point to the same campfires. The white box in \hri\ image (shown in c; also displayed in e) shows the location of the campfires that is presented in detail in Figure \ref{fig2}.
	}  \label{fig1}
\end{figure*} 

\section{DATA and Methods}\label{data}

In this study, we use L2 EUI data\footnote{\url{https://doi.org/10.24414/wvj6-nm32}} (calibrated images) from 20-May-2020 and 30-May-2020. The L2 data is the latest calibrated data product, which is suitable for scientific analysis. The observations of  these two days were taken when the instrument was in the commissioning phase. The HRI of EUI captured  images in 174 \AA\ channel (centered on the \FeIX\ and \FeX\ lines formed around 1 MK)  with high spatial resolution (pixel size of 0\arcsec.492; \citealt{rochus2020}) and high temporal cadence (5 to 10 s). All the campfires we studied here were detected in \hri\ 174 \AA. None of the campfires were visible in HRI$_{Lya}$. 

On 20-May-2020, the Solar Orbiter was located at a distance of 0.612 AU  from  the Sun, which means it observed the Sun from that  location  with a  pixel size  of 217 km (Figure \ref{fig1}a). 
As part of technical compression tests, \hri\ produced 174 \AA\ images between 21:20 and 22:17 UT, cycling through a 2 minute program that took 5 images at 10s cadence plus a 6th image 70 s later. These images have variable settings but 60 images are well exposed, un-binned and compressed at high quality levels. We find that campfires are visible in some binned data as well. Therefore, we have kept in the movies all the EUI frames including binned ones.

Whereas on 30-May-2020, the Solar Orbiter was positioned at a distance of 0.556 AU from the Sun (a pixel size of 198 km; Figure \ref{fig1}b) and it captured the \hri\  images for about 5 minutes at every 5 s. Figures \ref{fig1}a and \ref{fig1}b show the EUI/Full Sun Imager (FSI) 174 \AA\ images overlaid with HRI field of view, respectively, for 20-May-2020 and 30-May-2020. Both the images were created with  the use of JHelioviewer software \citep{muller17}.

In addition to EUI data, we used EUV images from SDO/AIA. AIA provides  full-disk solar images in seven different EUV channels. The AIA images have 0\arcsec.6 pixels and 12 s temporal cadence. For our analysis, we mainly used 304, 171, 193, and 211 \AA\ images, because we were able to recognize the EUI events clearly in these four channels. We found that campfires are rarely visible in hotter AIA channels (e.g. 335 and 94 \AA). We also checked that they are hardly visible in  AIA's UV channels (1600 and 1700  \AA).  To investigate the  photospheric magnetic field setting of campfires, we used line of sight magnetograms, of 45s series, from SDO/HMI. The HMI has 0\arcsec.5 pixels, 45 s temporal cadence and a noise level of about 7 G \citep{schou12,couvidat16}. Two consecutive magnetograms have been summed at each time step to enhance the visibility of weak field regions. 

We downloaded  two hours of AIA and HMI data for 20-May-2020 and an hour data for 30-May-2020, using JSOC data cutout service\footnote{\url{http://jsoc.stanford.edu/ajax/exportdata.html}}. The AIA and HMI data sets have been co-aligned using SolarSoft routines \citep{freeland98}. It is important to note that EUI and AIA observed the Sun from  different heliocentric distances. Solar Orbiter was at 0.612 AU and 0.556 AU from the Sun on 20-May-2020 and 30-May-2020, respectively. Therefore, the events would appear 3.22 minutes and 3.68 minutes earlier in EUI images than in the AIA images on 20-May-2020 and 30-May-2020, respectively. 
We have created movies, using EUI, AIA, and HMI data, to follow the evolution of campfires together with their underlying magnetic field. The movies and their corresponding figures display the same field of view. The contours, of $\pm$15 G, of HMI magnetograms are overplotted in the movies and figures. The Differential Emission Measure (DEM; \citealt{cheung15}) has been computed, using six AIA channels (171, 193, 211, 131, 335, and 94 \AA), to investigate the emission of campfires in different temperature bins.

We inspected the full-resolution \hri\ data and manually identified 52 campfires (from 44 different locations). We looked for those events that show similarities with the campfires in the Solar Orbiter/EUI press release images\footnote{\url{https://www.esa.int/Science_Exploration/Space_Science/Solar_Orbiter/Solar_Orbiter_s_first_images_reveal_campfires_on_the_Sun}}. After selecting the campfires in \hri\ 174 \AA, we looked for the same campfires in the AIA 171 \AA\ images. Then we co-aligned, using solar soft routines, the HMI line of sight magnetograms with respect to the AIA images, both onboard SDO. Therefore, the alignment between AIA images and HMI magnetograms is accurate in order to study the photospheric magnetic field of the campfires.

Our selected campfires are also consistent with the categories of campfires (loop-like, dot-like, and complex) in \cite{berghmans2021}. In addition, we find several campfires to be jet-like in their appearance.
All of our selected events are listed in Table \ref{tb: table}. We find that some of the campfires occur more than once from the same location.  We mainly obtained  the duration, length, and width of the campfires from the \hri\ images. We used AIA data to estimate the duration of those campfires that start/end a few minutes before/after the EUI-time coverage. 


In Table \ref{tb: table}, the measured length is the integrated distance of each campfire along their longer-extension, whereas the estimated width is the cross-sectional distance of the campfire. Both measurements are taken during the peak brightening of the campfires in \hri\ images. To estimate the uncertainty in the length and width, we repeated the measurements three  times at three nearby locations and then took average of the three measurements.  The listed uncertainty is the uncertainty of the mean (that is, standard deviation of the three measured lengths/widths from their mean, divided by the square root of 3). The  duration of campfires is the estimated time from when each of the campfire turns on until when it fades away in \hri\ images.

In Table \ref{tb: table} we have also listed the visibility of cool-plasma that is confirmed from the  AIA 304 \AA\ and/or 171 \AA\ images. We also mention the category of campfires in the last column -- the events appear as loop-like,  dot-like, jet-like, or complex  structures.

We created magnetic flux evolution plots to quantitatively assess the changes in magnetic flux as a function of time. We integrate flux inside a box covering those magnetic patches that are usually easier to isolate from the neighboring magnetic flux elements, and make sure no opposite flux flows across the boundary of the box. Furthermore, we also created HMI time-distance flux maps to present a detailed and clear picture of magnetic field evolution (e.g., Figure \ref{fig4}b).

\startlongtable
\begin{deluxetable*}{cccccccccc}
	\tablewidth{0pt}
	\tabletypesize{\footnotesize}
	\renewcommand{\arraystretch}{1.03}
	\setlength{\tabcolsep}{4.0pt} 
	\tablenum{1}
	\tablecaption{Measured Parameters of the Observed Campfires \label{tb: table}}
	\tablehead{
		\colhead{Event} & 
		\colhead{Time\tablenotemark{\scriptsize a}} & 
		\colhead{Location\tablenotemark{\scriptsize b}} & 
		\colhead{No. of\tablenotemark{\scriptsize c}} & 
		\colhead{Duration\tablenotemark{\scriptsize d}} &
		\colhead{Length\tablenotemark{\scriptsize e}} & 
		\colhead{Width\tablenotemark{\scriptsize f}}  & 
		\colhead{Visibility of\tablenotemark{\scriptsize g}} &
		\colhead{Discernible\tablenotemark{\scriptsize h}}  &
		\colhead{Category\tablenotemark{\scriptsize i}}  \\
		\colhead{No.} & 
		\colhead{(UT)} &
		\colhead{(x,y)arcsec} & 
		\colhead{CFs} &
		\colhead{(minutes)} &
		\colhead{(km)} &  
		\colhead{(km)} & 
		\colhead{cool plasma} & 
		\colhead{flux cancel.} &
		\colhead{of CFs} 
	}
	\startdata
	20-May-2020 \\
	\hline
	1  &  21:20:52   & -230,-270  & 3\tablenotemark{$\ast$}  & 2.1$\pm$10s   & 7100$\pm$330    &  1400$\pm$40    &  Y  & A\tablenotemark{j}  & loop-like  \\ 
	&  21:54:22   &    &   &  6.0 $\pm$12s\tablenotemark{\scriptsize $\star$}  & 7800$\pm$350     &  1350$\pm$100   &  Y  & A  &   \\ 
	2   &  21:26:12  & -235,-305  &  2  & 7.0$\pm$10s   & 5700$\pm$340    & 1500$\pm$160     & Y  & Y & loop-like \\ 
	&  21:58:22\tablenotemark{k}  &   &    & 4.0$\pm$12s\tablenotemark{\scriptsize $\star$}   & 6900$\pm$730    & 2700$\pm$900     & Y  & Y &  \\ 
	3  &  21:35:02  & -20,-250  &  3  &2.0$\pm$10s     &  7600$\pm$50   &  1600$\pm$360    &  Y  & A & loop-like \\ 
	&  21:44:52  &   &    & 1.5$\pm$10s   &  5500$\pm$150   &  1300$\pm$100    & Y  & A\\ 
	& 22:00:52   &   &    & 4.5$\pm$12s   &  9300$\pm$900   &  2100$\pm$300    & Y  & A\\ 
	4   &  21:48:22  & -580,380  &  2  & 7.0$\pm$10s    & 8700$\pm$550    &  1700$\pm$670    & Y  & Y  & complex \\
	&  22:10:32   &  &    & 2.0$\pm$10s    & 7000$\pm$640    &  1300$\pm$50    & Y  & Y \\
	5    &  21:26:52   & 285,55 &  1  & 17.0$\pm$20s    & 5500$\pm$920    &  2200$\pm$280    & Y  & Y & complex \\ 
	6    &  22:22:42  & 225,445  &  1  & 8.0$\pm$1m\tablenotemark{l}    &  9600$\pm$430   & 1900$\pm$220     &  Y  & Y & complex \\ 
	7   &  21:32:42   & -40,-370 &  2  & 6.0$\pm$10s    & 7300$\pm$350    &  2700$\pm$260    &  Y  & Y & complex\\
	&  22:08:12 &  &   & 129$\pm$24s\tablenotemark{l}&7000$\pm$550    & 2400$\pm$300 &  Y  & Y \\
	8  &  21:37:02   & -80,150  &  1   &17.0$\pm$10s    & 4900$\pm$60     &  1600$\pm$235    & Y  & Y   & loop-like  \\
	9   &  22:10:22   & -600,580  &  1  &12.0$\pm$30s\tablenotemark{l}        & 9500$\pm$280    &  2600$\pm$295  & Y  & Y & complex   \\
	10  &  22:13:02   & -530,570  & 1  &13.0$\pm$10s     &  10400$\pm$810   &  1900$\pm$780 & Y & Y \tablenotemark{m} & jet-like \\
	11  & 22:08:12    & -590,500  & 1  &8.0$\pm$10s     &  5000$\pm$160   &  2000$\pm$225    &Y & Y  & jet-like  \\
	12    & 22:04:52    &  -520,510  & 1  &26.0$\pm$20s     & 8100$\pm$340    &  1300$\pm$430    & Y & Y & dot-like \\
	13  & 22:11:02    & 280,43  & 1   &6.5$\pm$10s     & 6000$\pm$440    &  3800$\pm$225    & Y & Y  & jet-like\\
	14  &   22:16:22 & -45,-380  & 1 & 9.0$\pm$12s    & 2900$\pm$150    &  1400$\pm$250    &  Y  & Y & complex\\
	15  & 21:38:22 & -600,510  & 1 & 3.0$\pm$10s    & 3300$\pm$350    &  1500$\pm$240    &  Y  & Y & jet-like\\
	16  & 22:14:52 &  -580,365 &  1 & 8.0$\pm$10s    & 2700$\pm$350    &  1800$\pm$160    &  Y  & Y& loop-like \\
	17 & 22:08:22 & -605,395  &  2 & 5.5$\pm$10s    & 3700$\pm$100    &  1200$\pm$400    &  Y  & Y & jet-like \\
	& 22:14:32 &   &   & 2.0$\pm$10s    & 1300$\pm$150    &  1100$\pm$100    &  Y  & Y \\
	18 & 22:00:22 & -400,-20  &  1 & 1.0$\pm$10s    & 4900$\pm$570    &  2800$\pm$150    &  Y  & Y & jet-like\\
	19 & 22:16:22 & -400,0  &  1 & 8.0$\pm$12s\tablenotemark{\scriptsize $\star$}    & 2400$\pm$200    &  2100$\pm$130    &  Y  & Y\tablenotemark{m}  & dot-like\\
	20 & 22:11:22 & -380,0  &  1 & 4.0$\pm$10s    & 1000$\pm$10    &  960$\pm$40    &  Y  & Y\tablenotemark{m}  & dot-like\\
	\hline
	30-May-2020 \\
	\hline
	21   &  14:54:40     & -450,330  &  1  &  7.0$\pm$12s  &  9600$\pm$670   &  1100$\pm$315    & Y & Y\tablenotemark{n} & complex  \\ 
	22     & 14:54:30     &  -440,220  &  2\tablenotemark{$\ast$}  & 11.0$\pm$12s\tablenotemark{\scriptsize $\star$}   &  8400$\pm$85   &  1700$\pm$690    &  Y & Y & complex  \\ 
	23    & 14:57:50     & -335,145   &  1  & 4.0$\pm$5s   & 4700$\pm$185    &  1000$\pm$90 & Y & A\tablenotemark{o}  & complex  \\ 
	24    & 15:04:03     & -175,154   &  1  & 3.0$\pm$15s   & 7400$\pm$250    &  750$\pm$10    & Y & Y  & loop-like\\
	25      & 14:56:10    & -100,370  &  1  & 20.5$\pm$24s   & 5500$\pm$420    & 1600$\pm$370 & Y & Y    & complex \\ 
	26     & 14:55:50    & -305,350 & 3\tablenotemark{$\ast$} &  5.0$\pm$12s\tablenotemark{\scriptsize $\star$}&   4700$\pm$470  &  1000$\pm$220 &Y & Y  & jet-like   \\ 
	& 15:04:00    &  &   & 3.0$\pm$12s\tablenotemark{\scriptsize $\star$} & 5800$\pm$450 & 1200$\pm$100  &Y & Y    \\ 
	27   & 14:55:00     &  135,47  &  1  & 1.5$\pm$5s   &  4100$\pm$160   &  900$\pm$75    & N &  Y & loop-like \\
	28   & 14:55:55    & 20,-70 &  1  &  3.5$\pm$10s  &  4300$\pm$140   &  1000$\pm$120 & Y & Y & complex  \\
	29   & 14:54:00   & -190,50 &  1  & 3.5$\pm$5s   &  --\tablenotemark{p}    &  --    & Y\tablenotemark{q} & Y \tablenotemark{m} & complex \\
	30   &14:55:10   & -220,-100 &  1  & 198$\pm$10m\tablenotemark{\scriptsize $\star$}   &  6600$\pm$350   &  2000$\pm$200    & A\tablenotemark{q} &Y & complex  \\
	31   & 14:54:25    & 105,-180 &  1  & 22$\pm$12s   &  6700$\pm$1300   &  2100$\pm$340    & A & A  & complex \\
	32   & 14:57:30     & 265,-180 &  1  & 2.0$\pm$5s   &  3600$\pm$160   &   1200$\pm$35   & Y & Y  & complex \\
	33  &  14:54:45     & 95,-235  &   1 & 7.0$\pm$12s\tablenotemark{\scriptsize $\star$}   &  3600$\pm$135   & 1200$\pm$190     &   Y & A \tablenotemark{m} & jet-like  \\
	34   & 14:56:05      & -615,140  &  1  & 18$\pm$24s\tablenotemark{\scriptsize $\star$}   & 3500$\pm$55    &  1600$\pm$155    & A&Y  & complex \\
	35   & 14:57:20      & -575,170  &  3\tablenotemark{$\ast$} & 4.5$\pm$5s   &  6300$\pm$530   &   1100$\pm$155   &  A & Y & loop-like \\
	36   & 14:57:55      & 326,16  &   1 & 7.0$\pm$5s   &  3400$\pm$90   &  1200$\pm$130    & A&Y & complex  \\
	37   & 14:54:30      & -500,590  &   1 &  30$\pm$24s  & 7700$\pm$380    &  2700$\pm$250    & Y&Y  & jet-like \\
	38  &  14:55:10     &  -190,353 &   1 & 7.0$\pm$10s   &  6300$\pm$245   &  1800$\pm$35    & Y&Y\tablenotemark{m}   & complex \\
	39   &  14:56:50     &  -410,-235 &  1  & 3.0$\pm$5s   &  5300$\pm$285   &  1500$\pm$180    & A&Y & complex   \\
	40   &  14:54:25     &  -154,335 &  1  & 3.0$\pm$5s     &  3000$\pm$255   &  1000$\pm$140    &  A&A & dot-like \\
	41   &  14:57:15     &  -192,160 &  1  & 1.5$\pm$5s     &  2000$\pm$400   &  1150$\pm$270    &  Y&Y & complex  \\
	42   &  14:54:45     &  -148,320 &  1  & 1.0$\pm$5s     &  2100$\pm$140   &  700$\pm$100    &  A&A  & loop-like\\
	43   &  14:54:35     & -158,300  &  1  & 1.3$\pm$5s     &  1400$\pm$140   &  800$\pm$160    &  N &A  & dot-like \\
	44   &  14:57:20     & -195,-340  &  1  & 1.3$\pm$10s     &  750$\pm$20   &  700$\pm$70    &  -\tablenotemark{r} & -  & dot-like \\
	\noalign{\smallskip}\tableline \tableline \noalign{\smallskip} 
	average$\pm$1$\sigma$$_{ave}$  &   &     &     &  13.2$\pm$30 &  5400$\pm$2500     &   1600$\pm$640   &   \\
	\enddata
	
	\singlespace
	\tablecomments{	\\\textsuperscript{a}Approximate time of peak brightening in EUI 174 \AA\ images. 
		\\\textsuperscript{b}Approximate location of the campfires in EUI 174 \AA\ images.
		\\\textsuperscript{c}Total no. of campfires (CFs) from the same neutral line.
		\\\textsuperscript{d}Duration of the campfires in EUI 174 \AA\ images.
		\\\textsuperscript{e}Integrated length of the campfires during its peak brightening in  EUI 174 \AA\ images.
		\\\textsuperscript{f}Cross-sectional width of the campfires during its peak brightening in EUI 174 \AA\ images.
		\\\textsuperscript{g}Whether or not a cool-plasma structure is present at the base of the campfires in AIA 304 and/or 171 \AA\ images.
		\\\textsuperscript{h}Whether a discernible flux cancelation occurs at the base of the campfire.
		\\\textsuperscript{i}Category of the campfires.
		\\\textsuperscript{j}Canceling minority polarity flux is far from the PIL.
		\\\textsuperscript{k}EUI 174 \AA\ images are binned during the second event.
		\\\textsuperscript{l}Multiple structures brighten at the PIL even after the end of the campfire.
		\\\textsuperscript{m}Cancelation between `weak' flux elements.
		\\\textsuperscript{n}Flux coalescence can be seen at 14:53 UT in the negative flux clump during the event.
		\\\textsuperscript{o}Weak flux elements cancel 10 min before the rise of cool plasma.
		\\\textsuperscript{p}Ambiguous, there are many substructures making it difficult to estimate. 
		\\\textsuperscript{q}Cool plasma appears and disappears.
		\\\textsuperscript{r}Barely visible in AIA images.
		\\\textsuperscript{$\ast$}Only those campfires are characterized here that are covered by EUI (e.g. in Event No. 1, the third campfire is not covered by EUI, thus it is not characterized).
		\\\textsuperscript{$\star$}Duration estimated from AIA 171\AA\ images.
	}
\end{deluxetable*}


\begin{figure*}
	\centering
	\includegraphics[width=0.7\linewidth]{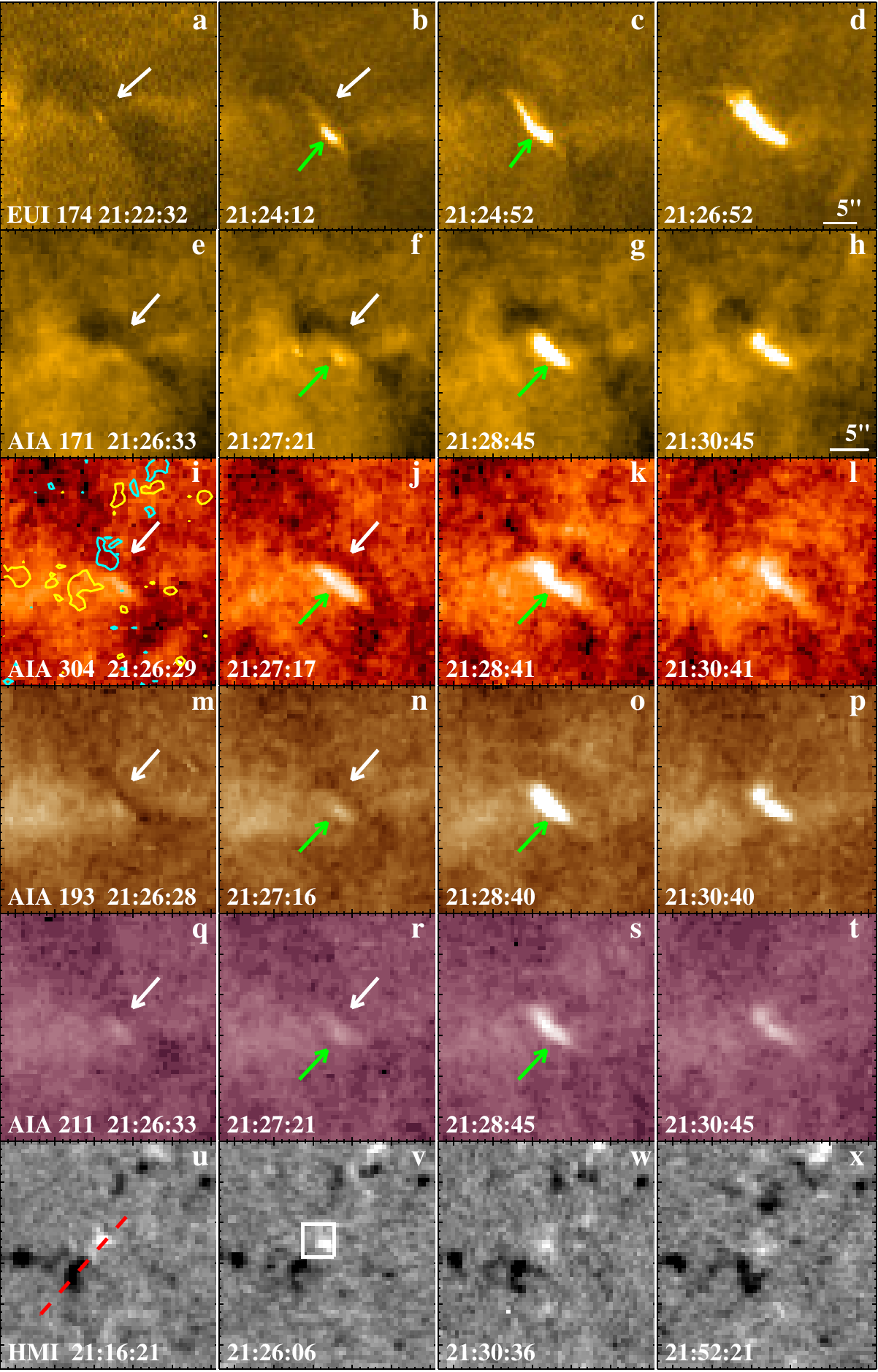}
	\caption{An example of a campfire observed on 20-May-2020. Panels (a-d) show 174  \AA\ \hri\ images of the campfire and display the same field of view as inside the white box of Figure \ref{fig1}c. A white horizontal bar in (d)  scales 5\arcsec\ distance for reference. For \hri\ on 20-May-2020 1\arcsec\ is equal to 443 km.
		Panels (e-h), (i-l), (m-p), and (q-t) show, respectively, the AIA 171, 304, 193, and 211 \AA\ images of the same campfire, and have the same field of view as in the white box of Figure \ref{fig1}e. Panels (u-x) show the HMI line of sight magnetograms of  the same region. The diagonal red dashed line in panel (u) show the location of the line along which the  time-distance map is created and shown in Figure  \ref{fig4}b. The white  box in (v)  shows  the area that  is  used  to calculate the positive magnetic flux plot shown in Figure \ref{fig4}a. The white  arrows point to  the  cool-plasma structure  and the green  arrows point to the  campfire itself. In panel (i), HMI contours, of levels $\pm$15 G (at 21:26:06) are overlaid, where cyan and yellow contours represent the positive and negative magnetic polarities, respectively. Note that EUI events appear 3.2 minutes earlier than in AIA images. The animation (Movie1-AIA) runs from 21:20 to 22:20 UT and the annotations and FOV are same as in this Figure, while the EUI animation (Movie1-So) runs from 21:20 to 21:32
UT and the animation is unannotated.
	}  \label{fig2}
\end{figure*} 

\section{Results}

\subsection{Overview}\label{over}

Figures \ref{fig1}c and \ref{fig1}d show some of the campfires observed by \hri\ on 20-May-2020 and 30-May-2020, respectively. Figures \ref{fig1}e and \ref{fig1}f  show the same campfires in AIA 171 \AA\ images. Some of the campfires that we have investigated in detail (listed in Table \ref{tb: table}) are marked by green arrows.

All these events appear in AIA 171 and 304 \AA\ images, but not as clearly as in \hri\ images (Figure \ref{fig1}).  Thus, we likely would not have noticed a few of these features in AIA 171 \AA\ images, without first having them seen in the higher-resolution EUI images. In Section \ref{cf1}, we present Campfire-2 of Table \ref{tb: table} in detail, and show six additional campfire examples, including their  DEMs, in Appendix (Section \ref{appen}).

\subsection{Presence of Cool Plasma at Campfire's Base}\label{cf1}
%

Figure \ref{fig2} shows an example of campfire observed on 20-May-2020. As noted in Table \ref{tb: table}, there are two campfires from the same location (Campfire-2 of Table \ref{tb: table}). Here, we present the first campfire from this location in detail. 


Figures \ref{fig2}(a-d) and \ref{fig2}(e-t) show the evolution of the campfire in \hri\ and AIA images, respectively. The campfire starts to brighten at 21:24:12 in \hri\ images (Figure \ref{fig2}a and Movie1-SO), reaching at its peak intensity at $\sim$21:26:12. The campfire lasts for about 7 minutes. Before and during the progression of the campfire, a small-scale cool-plasma structure appears at the same location of the campfire (see white arrow in Figure \ref{fig2}a,b). As the campfire grows with time, the cool plasma structure also expands upwards (from 21:24:12 to 21:25:02 in Movie1-SO) from the base of the campfire. The cool structure has a length of 8000$\pm$280 km  and a width of  1300$\pm$140 km  during the rise phase. 

The campfire and the cool-plasma structure are also visible in AIA 171, 304, 193, and 211 \AA\ images (Figure \ref{fig2}). However, in 193 and 211 \AA\ images the cool plasma is not as clearly visible as in 171 and 304 \AA\ images. We notice that the cool-plasma structure starts to rise at 21:24:57 in AIA 171 \AA\ (Movie1-AIA). After the rise of the  cool-plasma structure a brightening appears (at 21:26:57) turning into a campfire, at the location where  the cool-plasma structure was rooted prior to its rise (see Figures \ref{fig2}b,f). 

The campfire reaches its peak intensity at about 21:29:45  (Movie1-AIA). [Note: The campfire appears earlier in EUI images than in the  AIA images because of the different heliocentric distances of both the instruments from the Sun  (see  Section \ref{data}).] The cool plasma continues to rise and it erupts at 21:29:45. The appearance of the cool-plasma structure is very similar to the minifilaments that are seen to drive  typical coronal jets \citep[see, e.g.,][]{sterling15,panesar16b}. The average length and width of the cool structure (estimated using AIA 171 \AA\ images) are 11000$\pm$1500 km and 2300$\pm$100 km, respectively, which are also similar to the lengths and widths of  the pre-jet minifilaments \citep{panesar16b}.


Similarly, the majority of campfires are accompanied by a cool-plasma structure (Table \ref{tb: table}). See Section \ref{appen} for  six more examples of campfires (Figures \ref{fig1A}, \ref{fig2A}, \ref{fig3A}, \ref{fig4A}, \ref{fig5A}, and \ref{fig6A}). 

\begin{figure*}
	\centering
	\includegraphics[width=\linewidth]{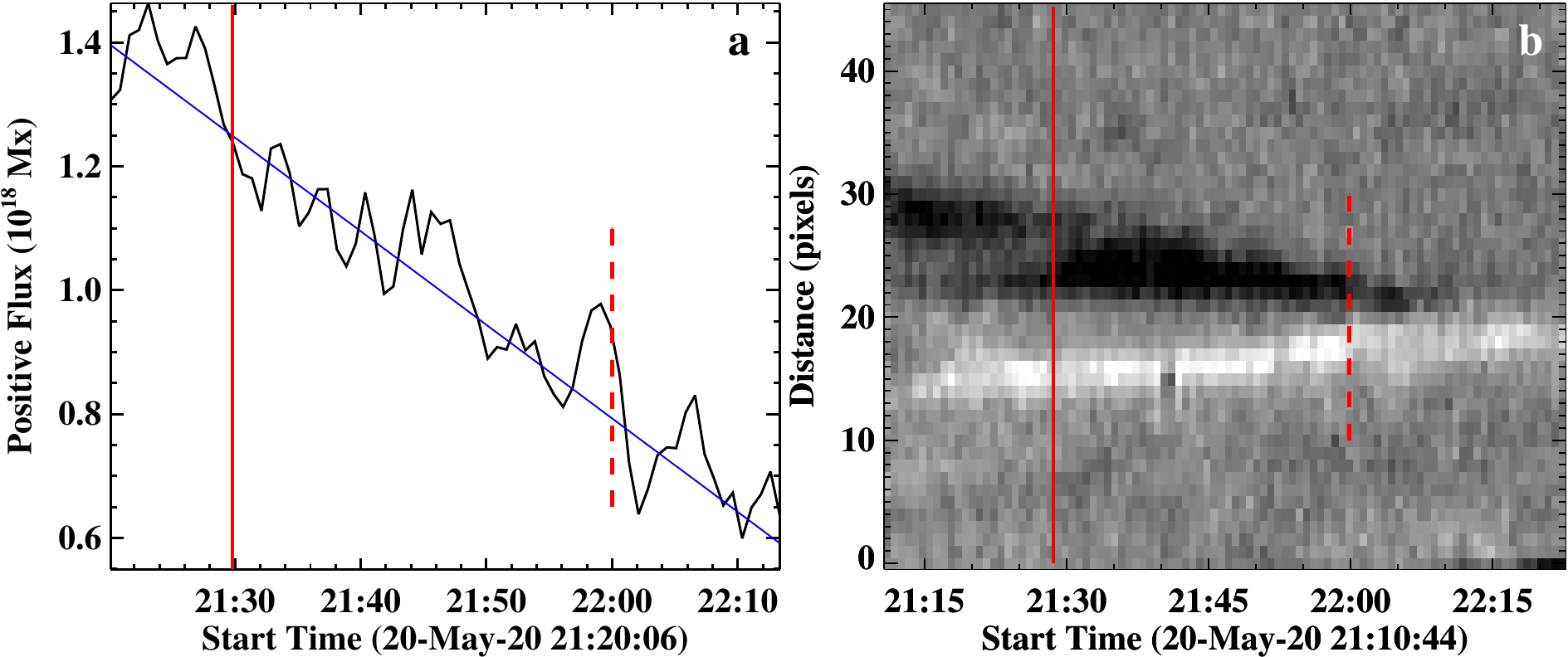}
	\caption{Magnetic field evolution at the base of the campfire presented in Figure \ref{fig2}. Panel (a) shows the integrated positive magnetic flux plot as a function of time computed inside the box shown in Figure \ref{fig2}v.  Panel (b)  shows the HMI time-distance maps along the red dashed lines in Figure \ref{fig2}u. The solid red lines show the peak time of the campfire. The red-dashed lines mark the time of the second campfire that occur at the same neutral line due to ongoing flux cancelation. The blue line is  the least-square fit of the flux evolution. 
	} \label{fig4}
\end{figure*} 

\subsection{Magnetic Field Evolution  at Campfire's Base}
Figures \ref{fig2}u-x display the photospheric line of sight magnetic field at the base of the campfire. The campfire occurs at the edge of the negative-polarity magnetic flux lane (Figure \ref{fig2}i). In this  region, the negative magnetic flux is in majority and positive flux is in minority.
 Both the campfire and cool-plasma reside above the same neutral line, between the majority-polarity magnetic flux patch (negative) and a merging minority-polarity flux patch (positive; Figure \ref{fig2}i).  We followed these flux patches for about two hours and noticed that there is flux cancelation going on between positive and negative flux patch  (Movie1-AIA and Figures \ref{fig2}u-x). 

 We carefully isolated the minority-polarity (positive) magnetic flux patch of the campfire-base  and  made a plot to show the magnetic flux evolution quantitatively (Figure \ref{fig4}a). The area of integrated flux is bounded by a white box in Figure \ref{fig2}v. Figure \ref{fig4}a shows that there is a  decrease in the positive magnetic flux, which is due to the continuous flux cancelation between the negative flux patch and a merging positive flux patch.  We interpret that the flux cancelation triggers the cool-plasma eruption and that cool-plasma drives the campfire. Another campfire appears at the same neutral line (see red-dashed line in Figure \ref{fig4}) apparently caused by the continuous flux cancelation. We estimate the average rate of flux decrease using the best-fit line in Figure \ref{fig4}a to be 1.0 $\times$ 10$^{18}$ Mx hr$^{-1}$. 
 
 Figure \ref{fig4}b shows the time-distance map along the red-dashed line of Figure \ref{fig2}u. The magnetic flux patches converge and cancel, and trigger the cool-plasma eruption accompanied by the campfire. 
 This scenario is similar to pre-jet minifilaments that are seen to form and erupt multiple times due to flux cancelation at the magnetic neutral line \citep{panesar17,sterling17,panesar18a}. Furthermore, the magnetic set-up of campfires are also similar to the magnetic environment of small-scale events (e.g. jetlets, explosive events, and coronal bright points) that are also seen to occur due to flux cancelation  \citep[e.g.,][]{panesar18b,panesar19,tiwari19,madjarska2019}.

Analogous to this example, we find that all the campfires, of Table \ref{tb: table}, reside above  magnetic neutral lines and most of them are accompanied with magnetic flux cancelation (Figures \ref{fig1A}, \ref{fig2A}, \ref{fig3A}, \ref{fig5A}, and \ref{fig6A}).
%

 \begin{figure*}
 	\centering
 	\includegraphics[width=\linewidth]{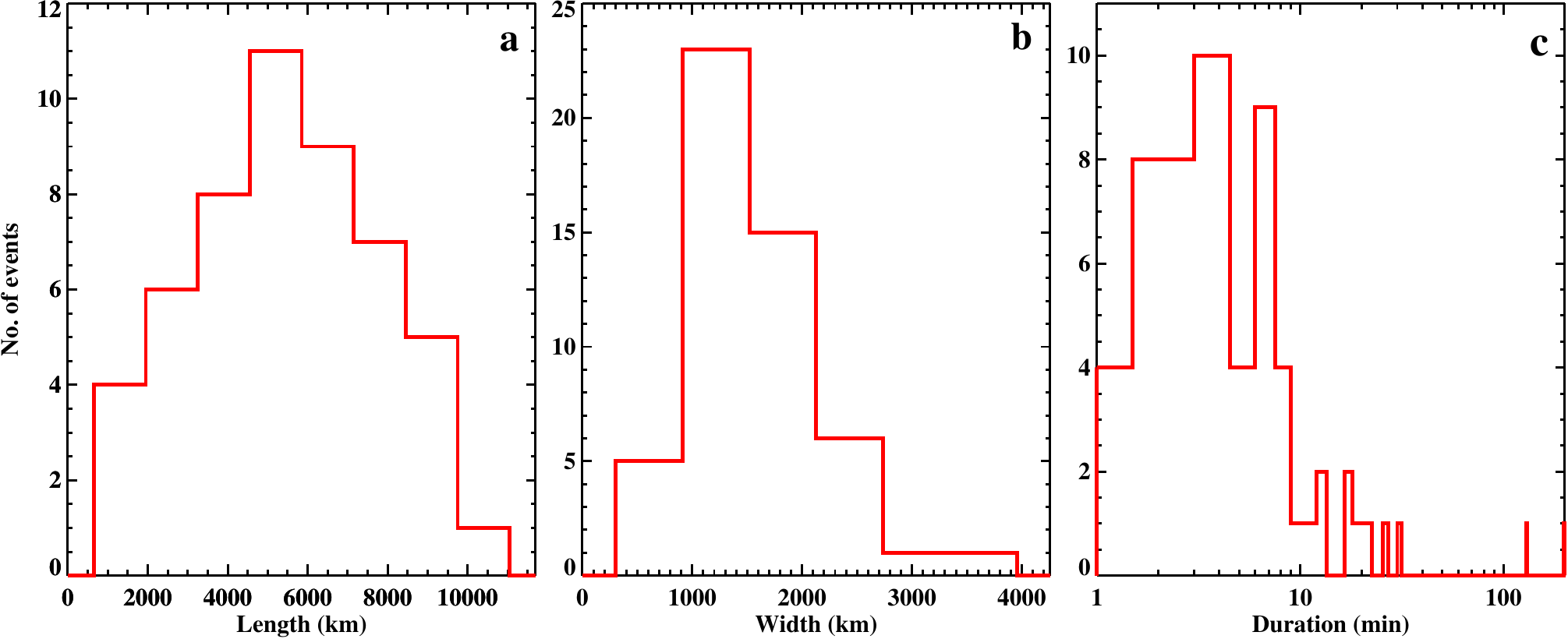}
 	\caption{Histograms of the physical parameters of 52 campfires listed in Table \ref{tb: table}. Panels (a), (b), and (c) show the histograms of the length, width, and duration of the campfires, respectively.  In panel (c) the x-axis is displayed in logarithm scale. The mean (median) values of the length, width, and duration are 5400$\pm$2500 km (5500 km), 1600$\pm$640 km (1500 km), and 13.2$\pm$30 minutes (6 minutes), respectively.  
 	}  \label{fig5}
 \end{figure*} 

\subsection{DEM}\label{dem}
We use images from six EUV AIA channels (171, 193, 211, 131, 335, and 94 \AA) to compute the DEM distributions of our campfires.  Figures \ref{fig7A}a,b,c  show the  total EM of the campfire region presented in Figure \ref{fig2} (campfire-2 of Table \ref{tb: table}) that  is  contained within the temperature range of log$_{10}$ T [5.7,6.0], [6.1,6.4], and [6.5,6.8], respectively. Evidently the campfire contains high coronal temperatures. Most of the EM lies in the log$_{10}$ T bins of 6 to 6.6 ($\sim$  1 to 4 MK). This particular campfire shows EM signals for low to high  temperature ranges (from log T$_{10}$  = [5.7,6.0] to [6.5,6.8]).

Figures  \ref{fig7A}(d-f) and \ref{fig7A}(g-i), show the DEM maps for campfires-8 and-13, from 20-May-2020. Both of these campfires are nicely visible in the temperature range of $\sim$ 0.5 to 2.5 MK but these do not show EM in 3 MK temperatures or above (in Figures \ref{fig7A}f,i). 

Similarly, Figure \ref{fig8A} shows the DEM distributions of campfires shown in Figures \ref{fig3A}, \ref{fig4A}, \ref{fig5A}, and \ref{fig6A}. All these campfires have large EM in the temperature range of $\sim$ 0.5 to 2.5 MK. Overall, all of these campfires appear to have emission in coronal temperatures -- this is consistent with the findings of \cite{berghmans2021}.

\subsection{Statistical Physical Properties}

Different physical properties of 52 campfires are listed in Table \ref{tb: table}. We find that all the campfires reside above the magnetic neutral lines. Most of the campfires (40 out of 52; exceptions are indicated in Table  \ref{tb: table}) are accompanied by  a  cool-plasma structure. Only in 2 events we did not find any evidence of cool-plasma -- these events are relatively smaller.  In 40 out of 52 events, we observe discernible flux cancelation. In a few cases, which are listed as ambiguous, campfires reside above very weak flux patches -- even in those cases HMI movies show visible flux cancelation before and/or during the events. We find that in majority of events, the cool-plasma structures are formed and triggered by flux cancelation. 

In Figures \ref{fig5}(a,b,c), we show histograms of the length, width, and duration of the 52 campfires (of Table \ref{tb: table}), respectively. Evidently, most of the campfires have lengths between 4000 to 6000 km, but they can be as small as 750 km and as long as 10400 km. The width of the most campfires ranges between 1000 to 1600 km, but they can be as narrow as 700 km and as wide as 3800 km. The sizes of these events are evidently larger than the spatial resolution of the \hri\ data ($\sim$400 km). The average  lengths of the campfires are smaller than the average lengths of Hi-C 2.1 jet-like events \citep{panesar20} and EUI microjets \citep{hou21}. The sizes of dot-like campfires are similar or smaller than the sizes of Hi-C 2.1 dot-like events \citep{tiwari19}. 
The majority of campfires (40) have durations of less than 10 minutes. But they can live as long as 198 minutes (Figure \ref{fig5}).


We estimated free magnetic energy (B$^2$$\times$V/8$\pi$) of the campfires based on their sizes and approximated coronal magnetic field of 20 G, and found them to range in the order of 10$^{26}$ to 10$^{27}$ erg. For estimating the volumes of dot-like campfires, we considered spherical geometry (V=4 $\pi r^3$/3), and took the average of their length and width as radius r.
Whereas for the loop-like, jet-like and complex campfires, we assumed a cylindrical geometry (V=$\pi r^2$h), and took lengths of the campfires as height h and their widths as radius r.
In the absence of true measured coronal magnetic field, and due to the fact that we use the projected lengths and widths of campfires, our evaluated free magnetic energies are only crude estimates.

\section{Discussion}

We have examined the evolution of 52 solar campfires using \hri\  images from Solar Orbiter and EUV images from \sdo/AIA, and investigated their magnetic origin using line of sight magnetograms from \sdo/HMI. We find that (i) all the campfires are located above magnetic neutral lines; (ii) in the majority of instances (79\%), campfires are accompanied by a cool-plasma structure; (iii) 77\% of campfires come from the sites of magnetic flux cancelation, analogous to coronal jets and jetlets; (iv) the cool-plasma structures are anchored at sites of flux cancelation, reminiscent of pre-jet minifilaments; (v) the campfires appear at coronal temperatures  ($\sim$ 0.5 to 2.5 MK);  (vi) their estimated free magnetic energies are in the order of 10$^{26}$ to 10$^{27}$ erg; and (vii) while many of the campfires are complex in their structure (previously unexplored), most of  them appear as a small-scale loop, a dot, or a coronal jet. In the following we discuss our findings:

\noindent \textit{Magnetic flux cancelation:} We find that all our campfires (in Table \ref{tb: table}) are rooted at the edges of  photospheric magnetic network flux lanes. Most of them occur, at neutral lines, at evident sites of magnetic flux cancelation (40 out of 52 show clear flux cancelation). In a few cases (listed as `ambiguous' in the second last column of Table \ref{tb: table}) (a) either magnetic flux cancelation occurs  between the weak flux patches, or (b) the canceling minority flux is a few pixels away from the base of the campfire. However, the minority-polarity flux patches are present at or around the base of each campfire, suggestive of magnetic flux cancelation being involved in generating all campfires.  

The flux cancelation rate of $\sim$10$^{18}$ Mx hr$^{-1}$, at campfire's base, are similar to the flux cancelation rates found for coronal jets in the quiet Sun regions and coronal holes (10$^{18}$ Mx hr$^{-1}$; \cite{panesar16b,panesar18a}),  and in large  penumbral jets (10$^{18}$ Mx hr$^{-1}$; \cite{tiwari18}, but are lower than the flux cancelation rates found for active region jets (10$^{19}$ Mx hr$^{-1}$; \cite{sterling17}), and surges in the core of active regions (10$^{19}$ Mx hr$^{-1}$ ;\cite{tiwari19}). 

\noindent \textit{Presence of cool-plasma:}  Most of the campfires (41 out of 52; exceptions are indicated in Table  1) are accompanied by  a  cool-plasma structure, which is present along the neutral line, at the base of the campfire. Only in 2 events we did not find any evidence of cool plasma. These events are relatively smaller -- therefore, the cool plasma might not be discernible. Furthermore, in some cases the campfires occur multiple times from the same neutral line, similar to jets, jetlets, and network jets \citep{raouafi14,tian14,panesar16b,panesar20b}, and each time they are accompanied with a cool-plasma structure. 

We infer that the presence of cool plasma structure might be an indicator of the presence of a magnetic flux rope, probably created by flux cancelation in the same way as observed for pre-jet minifilaments \citep{panesar17} and for typical solar filaments \citep{balle89,moore92}. We conjecture that the visible flux cancelation is a result of the submergence of the lower reconnected loops into the photosphere \citep{balle89,priest21,syntelis21}.

Alternatively, magnetic reconnection can directly lead to generation of campfires (without flux ropes) as proposed in some analytical theories and numerical simulations for coronal jets \citep[e.g.][]{yokoyama95,shibata11}. This could be the case for a few campfires that do not show clear presence of a cool plasma structure marked as `ambiguous/no' in the third last column of Table \ref{tb: table}). However, there are some simulations that do show flux rope formation in coronal jets \citep[e.g.][]{wyper18b,doyle19}.


\noindent \textit{Free magnetic energy:}    Our estimated free magnetic energies for campfires  (10$^{26}$ to 10$^{27}$  erg, as discussed earlier) are in agreement with that estimated for campfires in a numerical model by \cite{chen21}. Our estimated free magnetic energies are also in the range of that of  coronal hole jets \citep{pucci13} and coronal X-ray bright points  \citep[e.g.][]{priest94}. The campfires' free magnetic energies are about an order of magnitude lower in magnitude than the free magnetic energies of active region jets \citep{sterling17}, and of active region loops and sub-flares \citep{cirtain13,tiwari14}. Thus, campfires contain sufficient energy to heat the quiet-Sun corona, locally, which is consistent with their emission measure (EM) distributions. 

\noindent \textit{Similarities with other events:}  Based on their appearances in \hri\ images, we have categorized campfires as small loop-like, dot-like, jet-like, or complex structures (as listed in Table \ref{tb: table}). Our loop-like, jet-like/surge and dot-like events show similarities with the explosive events  in the core of an active region observed by Hi-C 2.1  \citep{tiwari19}. The Hi-C explosive events also appeared at sites of opposite-polarity magnetic flux patches, and are confined, thus are reminiscent to campfires, albeit them being located in a different magnetic environment. The loop-like campfires also look similar to the low-lying loop nanoflare events in the moss region observed by Hi-C \citep{winebarger13}.  The sizes of our jet-like campfires are  smaller than the sizes of Hi-C 2.1 jet-like events \citep{panesar19} and EUI microjets \citep{hou21}. Further, these two studies did not find any evidence of cool-plasma structure at the jet-base regions.

Some of the  campfires might have similarities with coronal X-ray bright points -- bright points are also seen to occur at the sites of magnetic flux cancelation \citep[e.g.][]{longcope98}. However, coronal bright points might be hotter  than campfires because they are  visible in hotter wavelengths e.g. in X-rays \citep{golub74} and/or in AIA 94 \AA\ \citep{madjarska2019}. Whereas our campfires are barely discernible in AIA 94 \AA.

Small-scale transient brightenings  in the quiet Sun are reportedly  located in the magnetic network lanes at the boundaries of supergranule cells \citep{porter91,falconer98,gosic14,attie16}, similar to our findings for campfires. Some of the campfires have similar sizes and lifetimes of explosive events found in the quiet Sun regions \citep{dere89}. However, the explosive events in the above-mentioned study are found to have lower temperatures and thus most probably are transition region events \citep[e.g.][]{chae98,innes13,gupta15,huang19}.

The physical properties of most campfires are in agreement with the above mentioned events (i.e., loops, dots, coronal jets and coronal bright points). Therefore the term `campfire' includes different variety  of small-scale solar features/events.


The presence of cool-plasma structure in most of our campfires  provides evidence that tiny flux ropes are plausibly present in many of solar features, everywhere on the Sun. The magnetic flux cancelation seems to build and  trigger tiny flux rope eruptions in campfires, similar to the  pre-jet minifilaments \citep{young14a,adams14,panesar16b,panesar17,panesar18a,mcglasson19,muglach21,shen21} and typical solar filaments  \citep[that drive CMEs e.g.][]{sterling18}.  The only major difference lies in that the cool-plasma eruptions  in jets occur along far-reaching field lines, whereas in campfires  the cool-plasma eruptions are mostly confined and seem to occur at the base of closed-field lines. That is plausibly why most campfires are not seen to erupt outwards.

Our results support the simulations of campfires \citep{chen21} -- these are at coronal temperature and show evidence of magnetic flux ropes. However, our observations show that the presence of cool structures is much more common than inferred in the simulation \citep{chen21} where only one out of the seven campfires showed a flux rope. 

Our observational results provide new insights into the understanding of coronal heating by small-scale campfires, and elucidate that cool plasma structures might accompany most solar eruptions (ejective or confined). These tiny flux ropes are most probably created by small-scale magnetic flux cancelation, plausibly driven by photospheric random convective flows.

\section{Conclusions}

We investigate the magnetic origin of solar campfires by using Solar Orbiter/EUI and \sdo\ (AIA and HMI) observations. Although many campfires are `complex', the term \textit{campfire} encompasses a wide variety of small-scale coronal events (e.g. small-scale loops, dots, and coronal jets).
They occur at the edges of  photospheric magnetic network flux lanes. Majority of the campfires are accompanied by a cool-plasma structure, which resides above the magnetic neutral line, where evidently magnetic flux cancelation takes place. Thus, campfires bear similarities with solar eruptions such as coronal jets and CME-producing filament eruptions. Our observations suggest that the presence of cool-plasma, i.e., magnetic flux ropes in the solar atmosphere, is more common than previously thought. The campfires contain coronal temperatures and plausibly are small-scale magnetic reconnection events.

\begin{acknowledgments}
We thank an anonymous referee for constructive comments. 
NKP acknowledges support from NASA’s SDO/AIA (NNG04EA00C) and HGI grant  (80NSSC20K0720).  SKT gratefully acknowledges support by NASA HGI grant (80NSSC21K0520) and NASA contract NNM07AA01C.
We acknowledge the use of  Solar Orbiter/EUI and  \sdo/AIA/HMI data. Solar Orbiter is a space mission of international collaboration between ESA and NASA, operated by ESA. The EUI instrument was built by CSL, IAS, MPS, MSSL/UCL, PMOD/WRC, ROB, LCF/IO with funding from the Belgian Federal Science Policy Office (BELSPO/PRODEX); the Centre National d’Etudes Spatiales (CNES); the UK Space Agency (UKSA); the Bundesministerium für Wirtschaft und Energie (BMWi) through the Deutsches Zentrum für Luft- und Raumfahrt (DLR); and the Swiss Space Office (SSO). AIA is an instrument onboard the Solar Dynamics Observatory, a mission for NASA’s Living With a Star program. This work has made use of NASA ADSABS.

\end{acknowledgments}

\bibliographystyle{aasjournal}


\section{Appendix information}\label{appen}

\begin{figure}[ht]
	\centering
	\includegraphics[width=\linewidth]{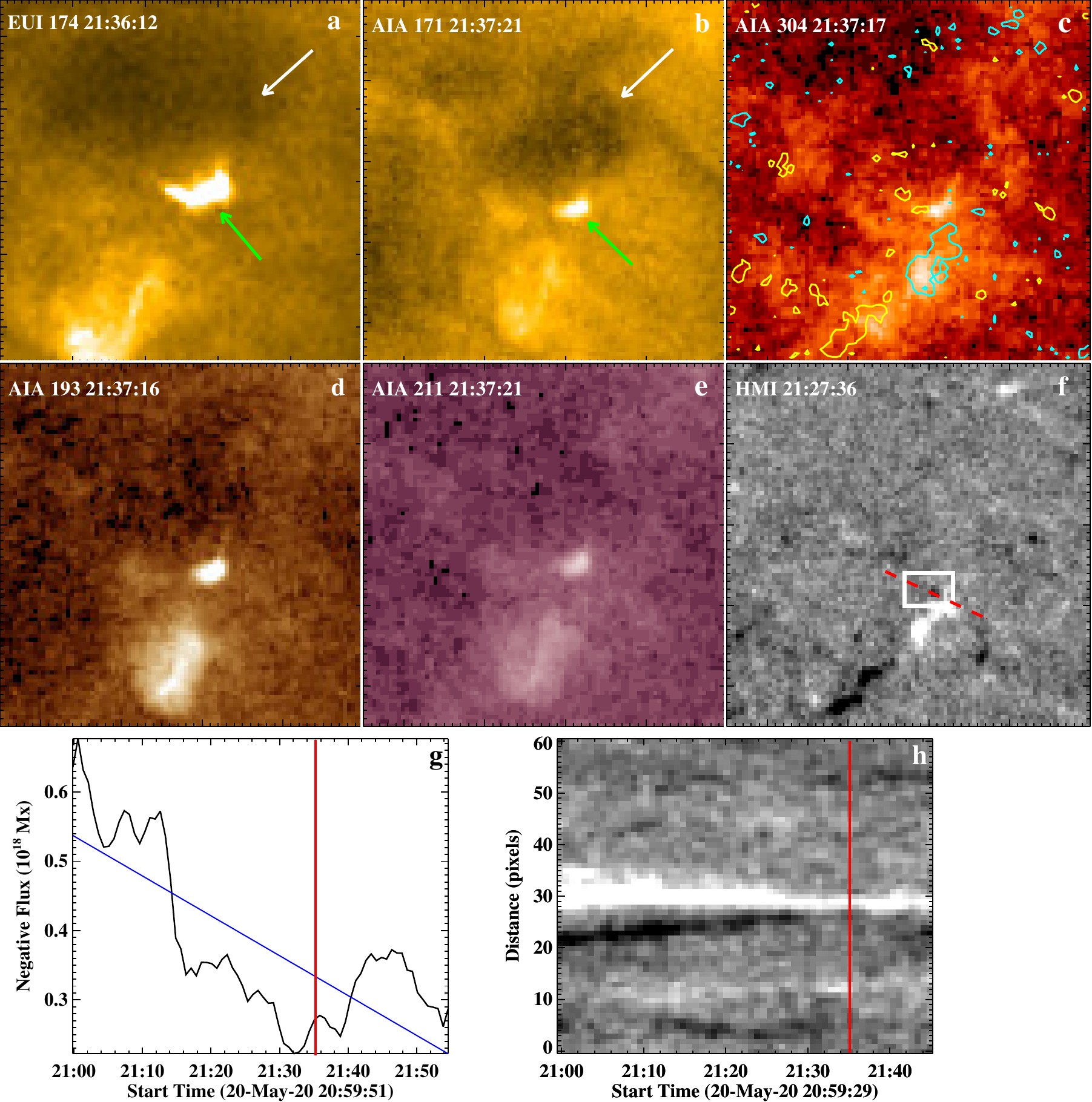}
	\caption{Campfire-8 observed on 20-May-2020 (Table \ref{tb: table}): Panel (a) shows 174  \AA\ \hri\ image of the campfire.  Panels (b), (c), (d), and (e) show the AIA 171, 304, 193, and 211 \AA\ images  of the campfire, respectively. The white  arrows point to  the  cool-plasma structure  and the green  arrows point to the  campfire. Panel (f) shows the HMI magnetogram of  the same region. The red  dashed line in (f) shows  cut for the  time-distance map in panel (h). The white box in (f)  shows  the area that  is  used  to calculate the negative magnetic flux plot shown in panel (g). Panels (g) and (h), respectively, show the negative flux plot versus time and the HMI time-distance map. The peak in (g) at 21:45 is due to flux coalescence. The rate of magnetic flux cancelation is 0.4 $\times$ 10$^{18}$ mx hr$^{-1}$. The red line marks the start time of the campfire. In panel (c), HMI contours, of levels $\pm$15 G, of 21:38:06 are overlaid, where cyan and yellow contours represents the positive and negative polarities, respectively.  The animation (Movie1A-AIA) runs from  21:00 to 22:08 UT and the annotations and FOV are same as in this Figure, while the EUI animation (Movie1A-So) runs from 21:32 to 21:49 UT and the animation is unannotated.
	} \label{fig1A}
\end{figure} 

\begin{figure}
	\centering
	\includegraphics[width=\linewidth]{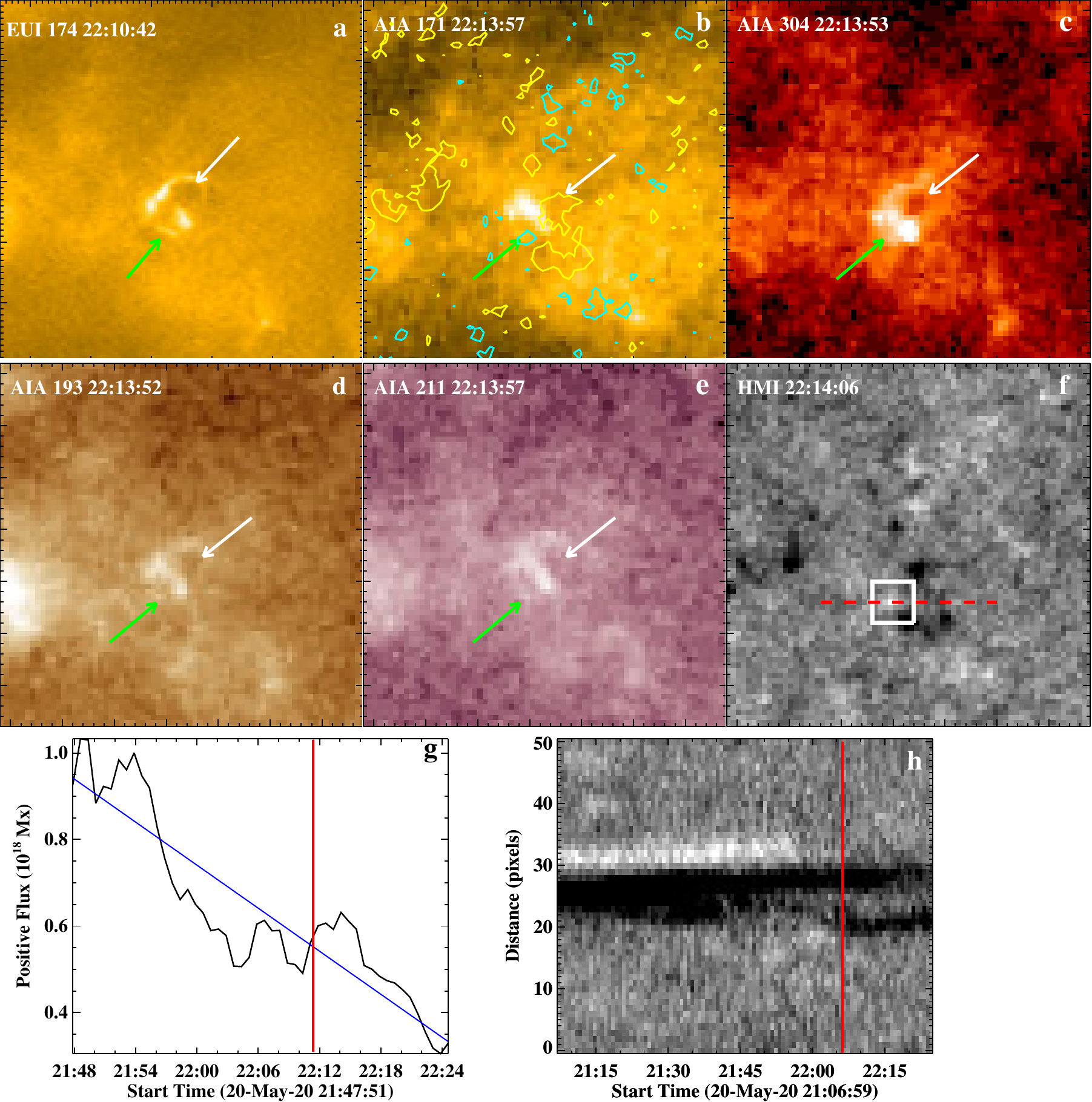}
	\caption{Campfire-13 observed on 20-May-2020 (Table \ref{tb: table}): Panel (a) shows the 174  \AA\ \hri\ image of the campfire.  Panels (b), (c), (d), and (e) show the AIA 171, 304, 193, and 211 \AA\ images  of the campfire, respectively. The white  arrows point to  the  cool-plasma structure  and the green  arrows point to the  campfire. Panel (f) shows the HMI magnetogram of  the same region. The white box in (f)  shows  the area that  is  used  to calculate the positive magnetic flux plot shown in panel (g). The red  dashed line in (f) shows east-west cut for the  time-distance map in panel (h). Panels (g) and (h) show the positive flux plot versus time and the HMI time-distance map, respectively. Flux cancelation rate is 1.0 $\times$ 10$^{18}$ mx hr$^{-1}$.  The red line marks the start time of the campfire in (g) and (h). In panel (b), HMI contours, of levels $\pm$15 G, of 21:14:06 are overlaid, where cyan and yellow contours represents the positive and negative magnetic polarities, respectively. The animation (Movie2A-AIA) runs from 21:47 to 22:24 UT and the annotations and FOV are same as in this Figure, while the EUI animation (Movie2A-So) runs from 22:08 to 22:17 UT and the animation is unannotated.
	} \label{fig2A}
\end{figure} 

\begin{figure}
	\centering
	\includegraphics[width=\linewidth]{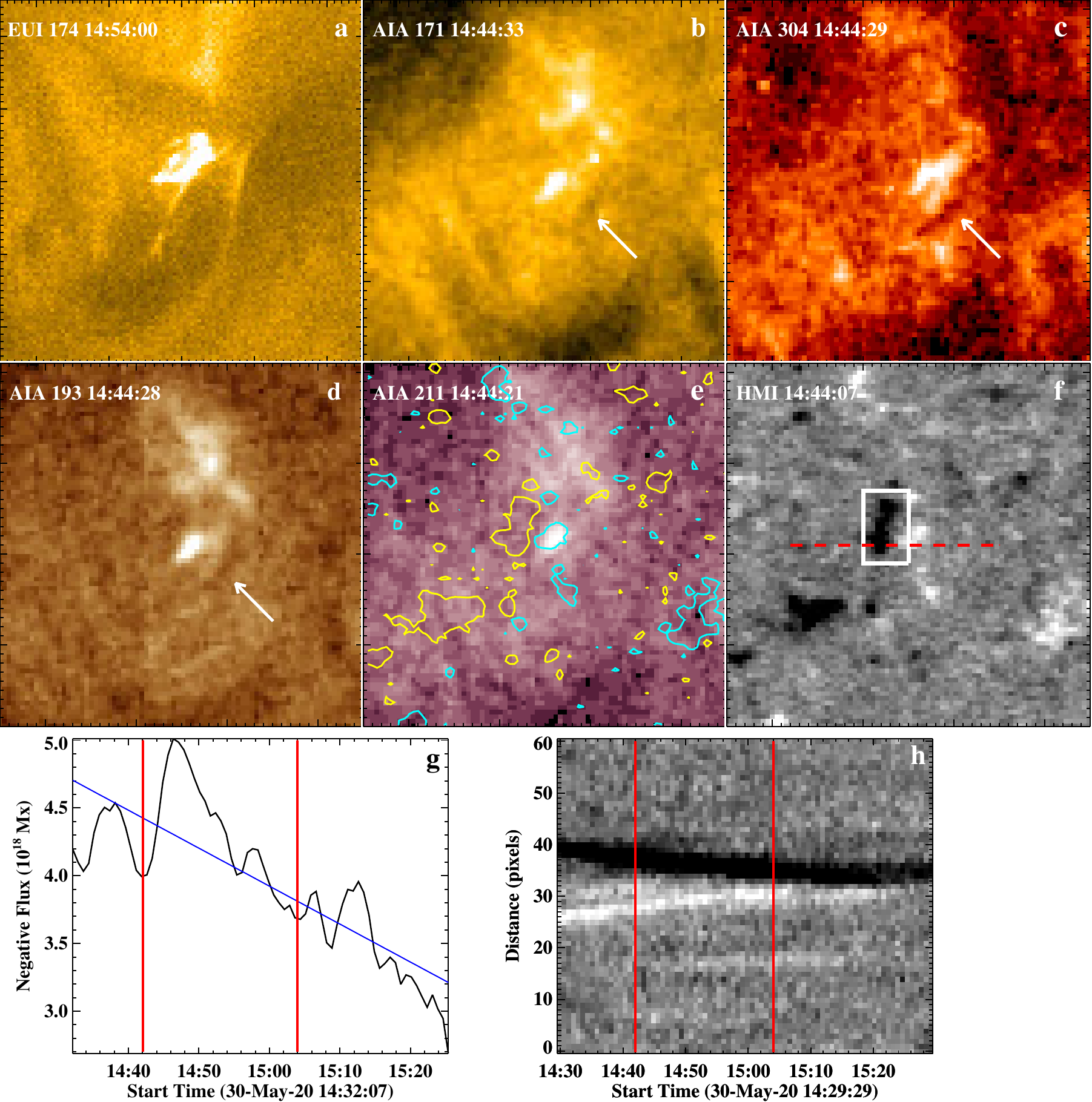}
	\caption{Campfire-22 observed on 30-May-2020 (Table \ref{tb: table}): Panel (a) shows the 174  \AA\ \hri\ image of the campfire.  Panels (b), (c), (d), and (e) show the AIA 171, 304, 193, and 211 \AA\ images  of the campfire, respectively. The white  arrows point to  the  cool-plasma structure  that is accompanied by the campfire. Panel (f) shows the HMI magnetogram of  the same region. The white box region in (f)  shows  the area that  is  used  to calculate the negative magnetic flux plot shown in panel (g). The red  dashed line in (f) shows east-west cut for the  time-distance map in panel (h). Panels (g) and (h) show, respectively, the negative flux plot versus time and the HMI time-distance map  along the red-dashed line in (f). The peak in Panel (g) at 14:48 is due to flux coalescence. Flux cancelation rate is 1.7 $\times$ 10$^{18}$ mx hr$^{-1}$.  The vertical red lines show the start time of both the campfires from the same location. In panel (e), HMI contours, of levels $\pm$15 G, of 14:44:07 are overlaid, where cyan and yellow contours represents the positive and negative magnetic polarities, respectively.  This campfire is same as that shown in Figure C5/C6 in \cite{berghmans2021}. The animation (Movie3A-AIA) runs from 14:30 to 15:28 UT and the annotations and FOV are same as in this Figure, while the EUI animation (Movie3A-So) runs from 14:54 to 15:04 UT and the animation is unannotated.
	} \label{fig3A}
\end{figure} 

\begin{figure}
	\centering
	\includegraphics[width=\linewidth]{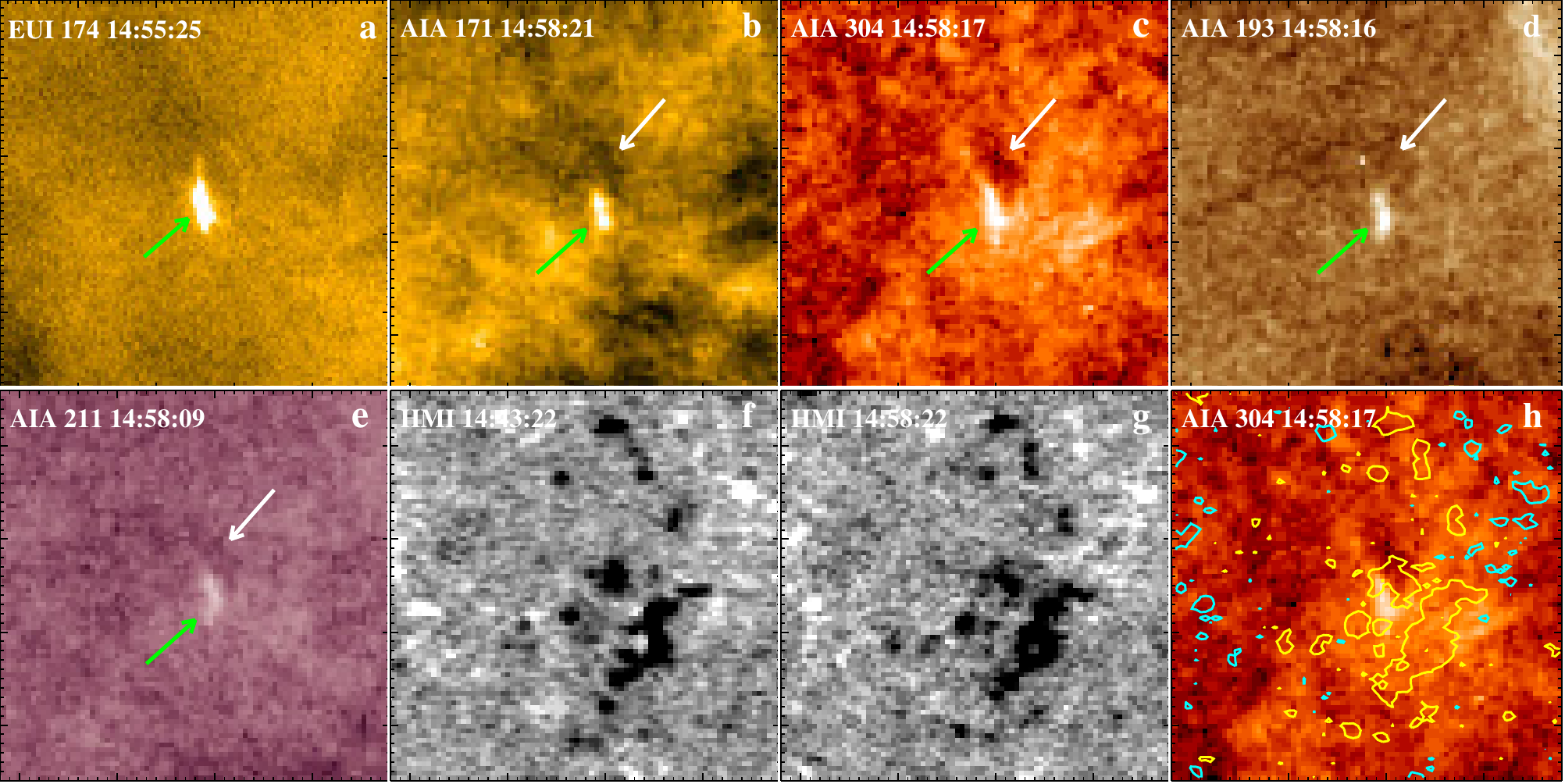}
	\caption{Campfire-23 observed on 30-May-2020 (Table \ref{tb: table}): Panel (a) shows the 174  \AA\ \hri\ image of the campfire.  Panels (b), (c), (d), and (e) show the AIA 171, 304, 193, and 211 \AA\ images  of the campfire, respectively. The white  arrows point to  the  cool-plasma structure  that is accompanied by the campfire. The green arrows show the campfire. Panels (f) and (g) show the HMI magnetograms of  the same region. Panel (f) shows the weak magnetic flux patches that cancel 10 min before the rise of cool plasma.
		In panel (e), HMI contours, of levels $\pm$15 G, of 14:58:22 are overlaid on AIA 304 \AA\ image, where cyan and yellow contours represents the positive and negative magnetic polarities, respectively. The animation (Movie4A-AIA) runs from 14:35 to 15:09 UT and the annotations and FOV are same as in this Figure, while the EUI animation (Movie4A-So) runs from 14:54 to 15:04 UT and the animation is unannotated.
	 } \label{fig4A}
\end{figure} 

\begin{figure}
	\centering
	\includegraphics[width=\linewidth]{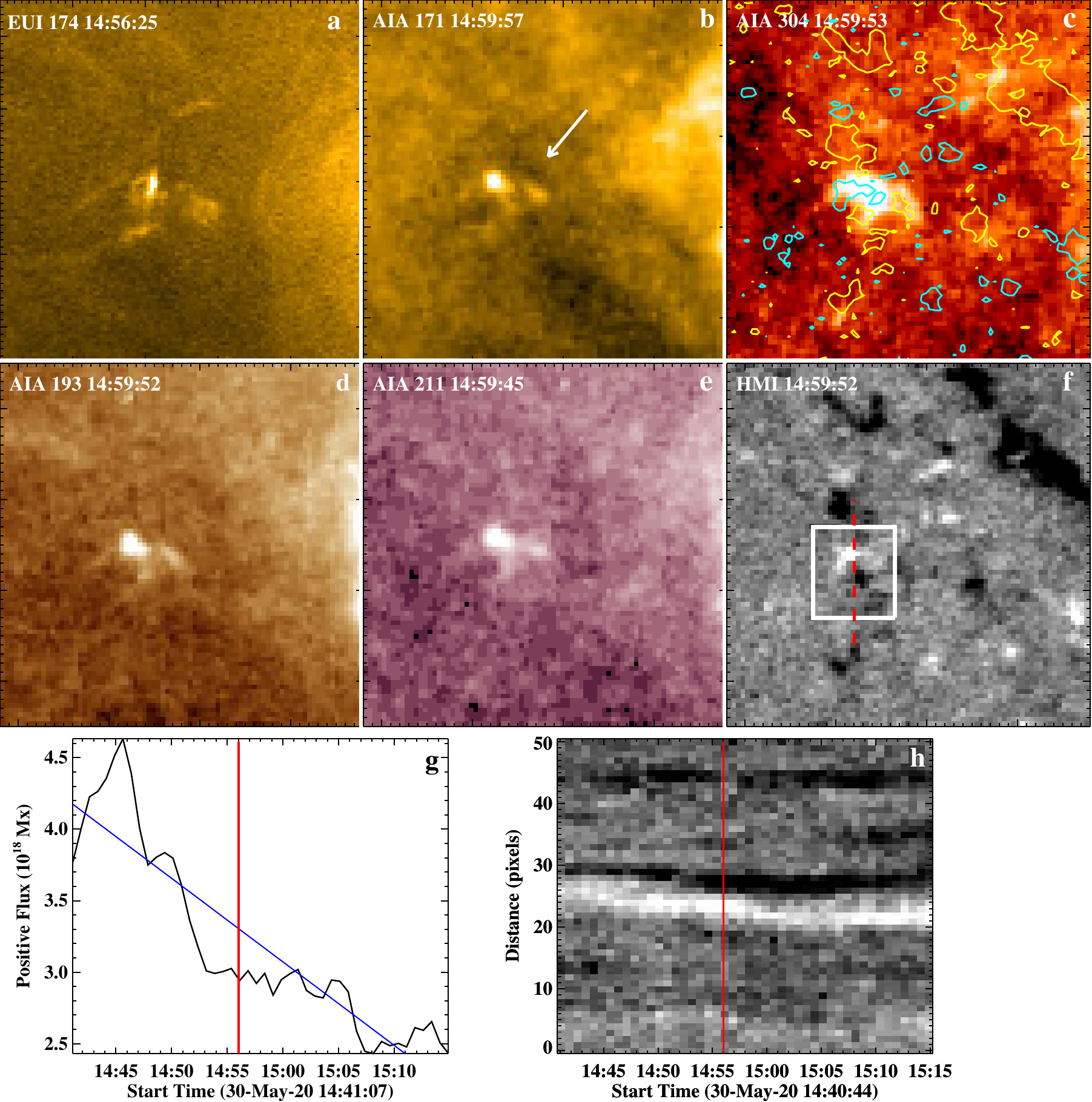}
	\caption{Campfire-25 observed on 30-May-2020 (Table \ref{tb: table}): Panel (a) shows the 174  \AA\ \hri\ image of the campfire.  Panels (b), (c), (d), and (e) show the AIA 171, 304, 193, and 211 \AA\ images  of the campfire, respectively. The white  arrow in (b) points to  the possible structure of the  cool-plasma; the cool-plasma is not visible in any other AIA channel. Panel (f) shows the HMI magnetogram of  the same region. The white box in (f)  shows  the area that  is  used  to calculate the positive magnetic flux plot shown in panel (g). The red  dashed line in (f) shows north-south cut for the  time-distance map in panel (h). Panels (g) and (h) shows the positive flux plot versus time and the HMI time-distance map. Flux cancelation rate is 3.6 $\times$ 10$^{18}$ mx hr$^{-1}$. The red line marks the start time of the campfire in (g) and (h). In panel (c), HMI contours, of levels $\pm$15 G, of 14:59:52 are overlaid, where cyan and yellow contours represents the positive and negative magnetic polarities, respectively.  The animation (Movie5A-AIA) runs from 14:41 to 15:17 UT and the annotations and FOV are same as in this Figure, while the EUI animation (Movie5A-So) runs from 14:54 to 15:04 UT and the animation is unannotated.
	} \label{fig5A}
\end{figure} 

\begin{figure}
	\centering
	\includegraphics[width=\linewidth]{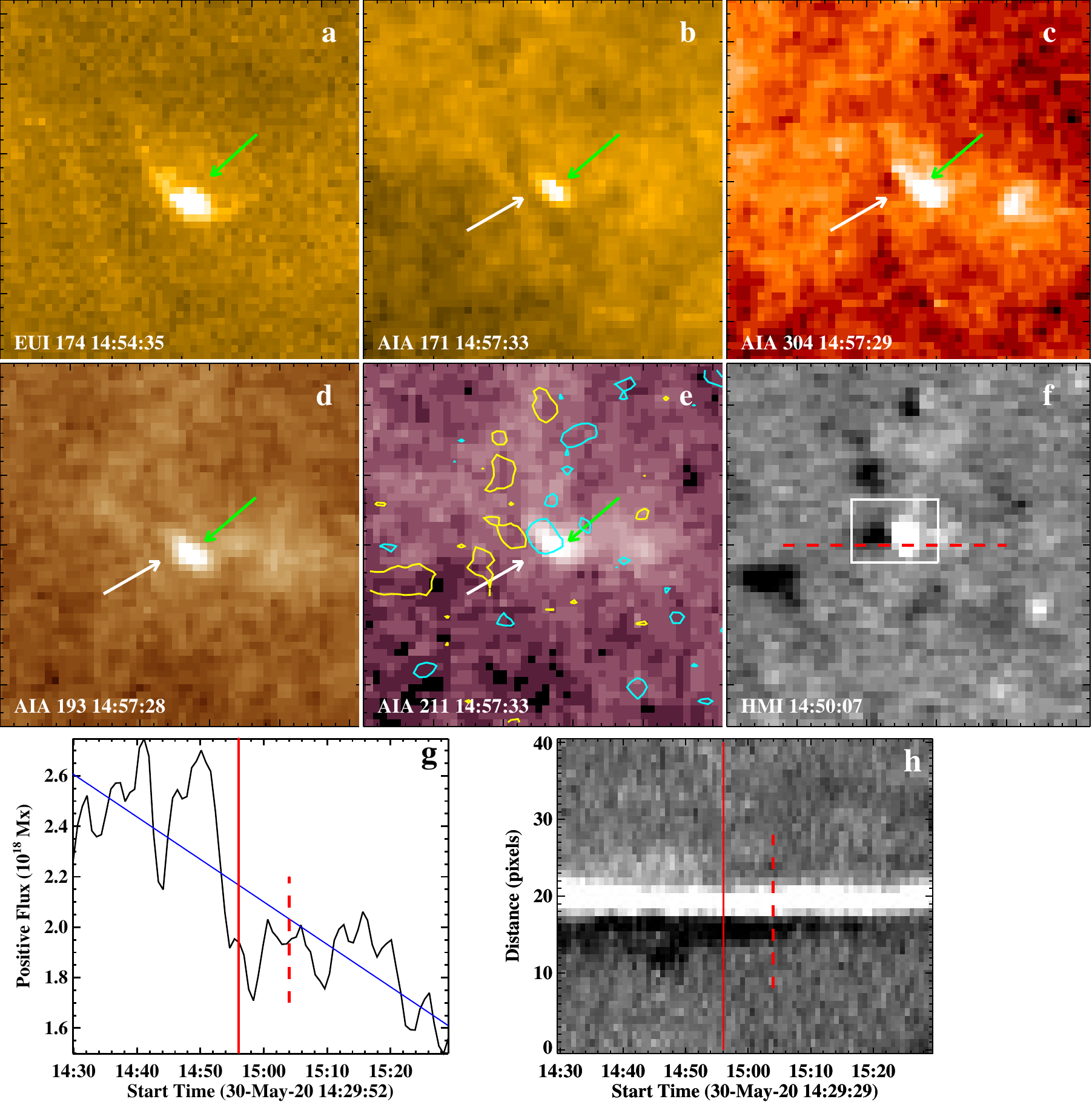}
	\caption{Campfire-26 observed on 30-May-2020 (Table \ref{tb: table}): Panel (a) shows the 174  \AA\ \hri\ images of  the campfire.  Panels (b), (c), (d), and (e) show the AIA 171, 304, 193, and 211 \AA\ images  of the same campfire, respectively. The white and green arrows point to  the cool-plasma structure and campfire, respectively. Panel (f) shows the HMI magnetogram of  the same region. The white box in (f)  shows  the area that  is  used  to calculate the positive magnetic flux plot shown in panel (g). The red  dashed line in (f) shows east-west cut for the  time-distance map in panel (h). Panels (g) and (h) shows the positive flux plot versus time and the HMI time-distance map, respectively. Flux cancelation rate is 0.95 $\times$ 10$^{18}$ mx hr$^{-1}$.  The solid red lines show the peak time of the first campfire. The red-dashed lines mark the times of the second campfire that occur at the same neutral line due to ongoing flux cancelation. The blue line in (g) is  the least-square fit of the flux evolution. In panel (e), HMI contours, of levels $\pm$15 G, of 14:56:52 are overlaid, where cyan and yellow contours represents the positive and negative magnetic polarities, respectively.  The animation (Movie6A-AIA) runs from 14:30 to 15:29 UT and the annotations and FOV are same as in this Figure, while the EUI animation (Movie6A-So) runs from 14:54 to 15:04 UT and the animation is unannotated.
	} \label{fig6A}
\end{figure} 

\begin{figure}
	\centering
	\includegraphics[width=\linewidth]{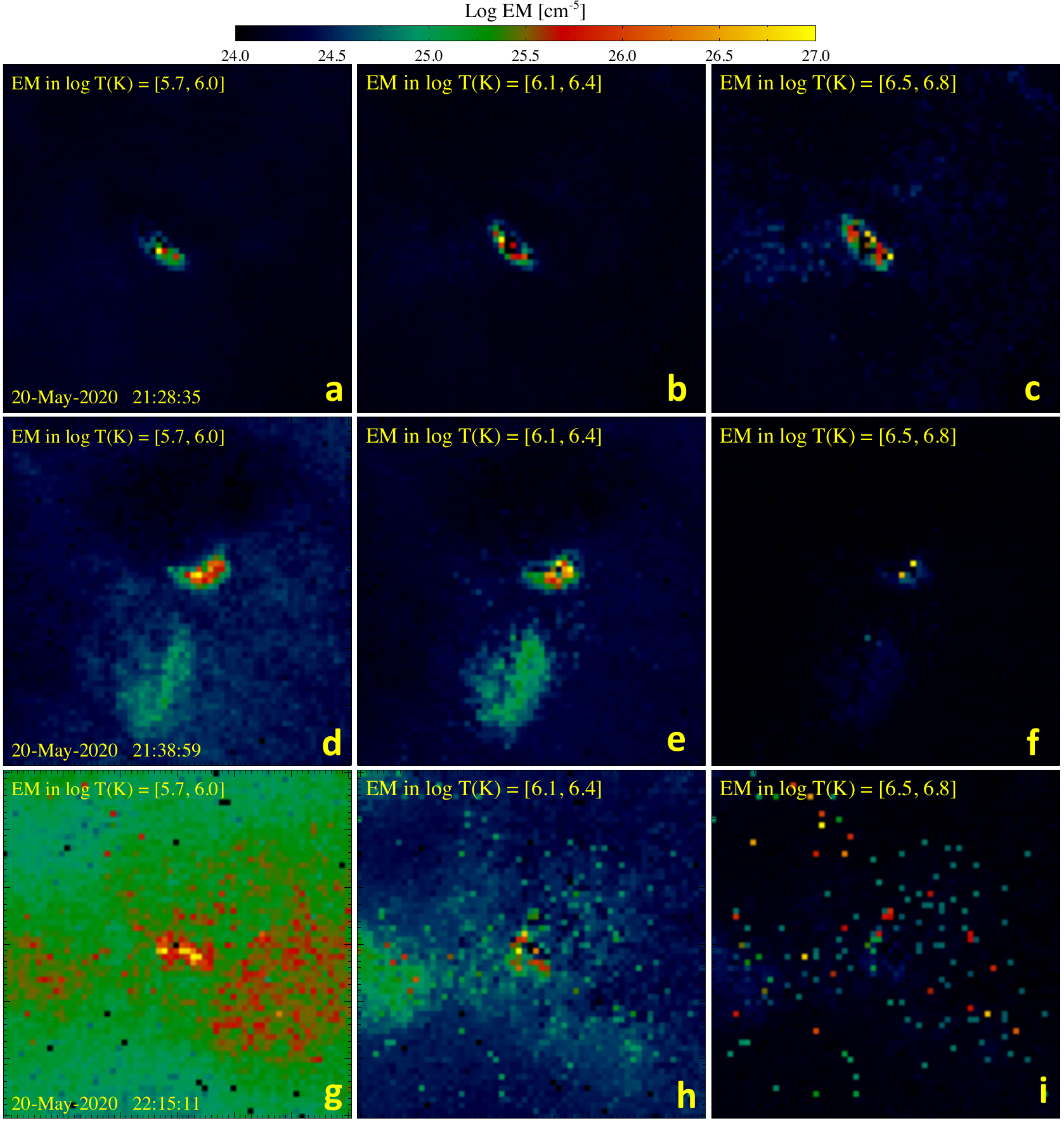}
	\caption{DEM distribution of campfires observed on 20-May-2020: Panels (a-c), (d-f), and (g-i) display the DEM maps of campfires shown in Figures \ref{fig2}, \ref{fig1A}, and  \ref{fig2A}, respectively.
		The EM for each campfire is displayed in three different temperature bins. The  color-bar on the top of the image indicates  the  total  EM  included within each log$_{10}$ T range mentioned in the upper left corner of each panel. The DEM maps show that these campfires have significant EM at coronal temperatures (a,b,c,d,e,g,h). However, in panels (f) and (i) there is barely any emission at temperatures above log$_{10}$ T(K) = 6.5.} \label{fig7A}
\end{figure} 

\begin{figure}
	\centering
	\includegraphics[width=0.9\linewidth]{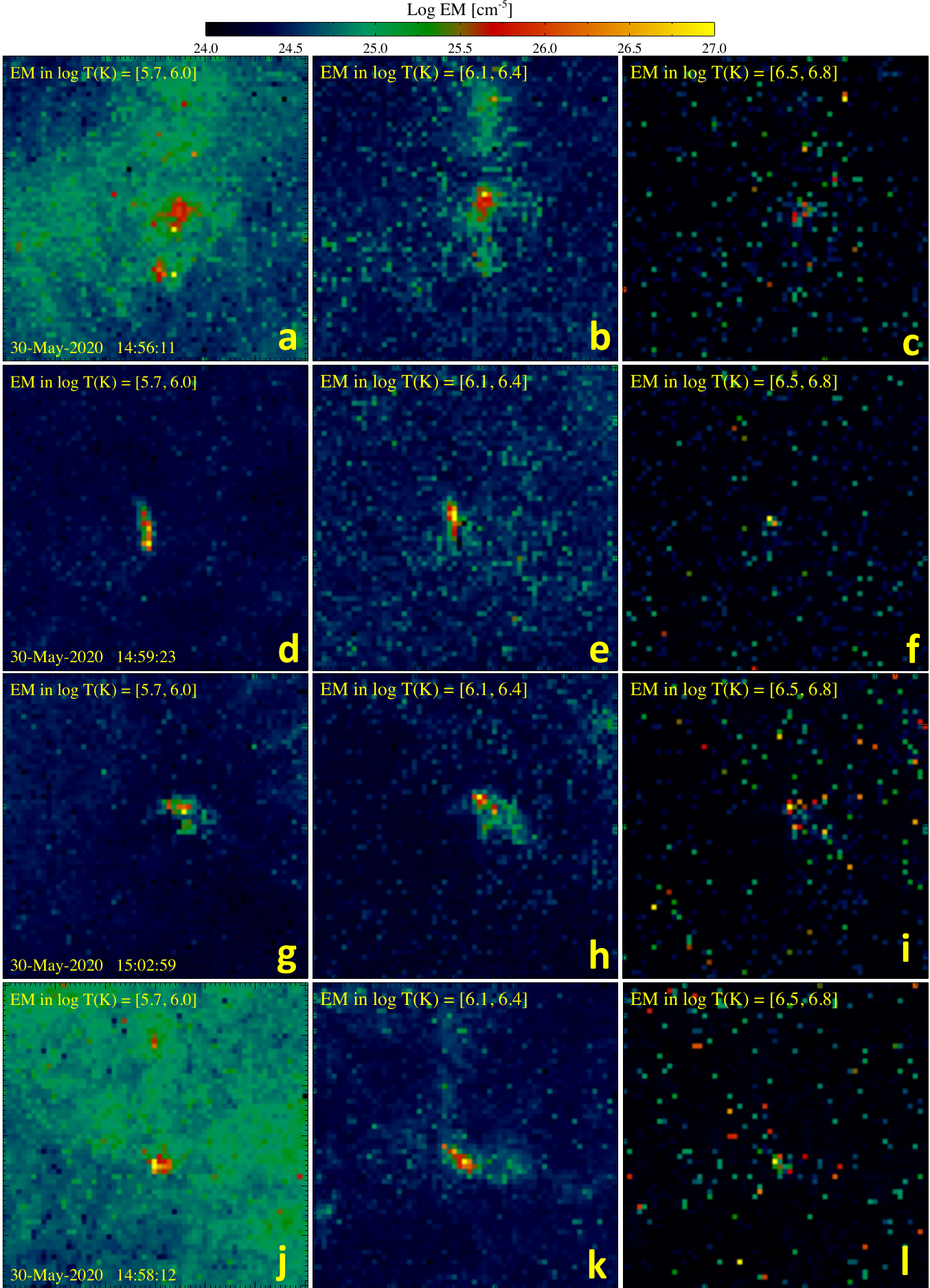}
	\caption{DEM distribution of campfires observed on 30-May-2020: Panels (a-c), (d-f), (g-i),  and (j-l) show the DEM maps of campfires shown in Figures \ref{fig3A}, \ref{fig4A}, \ref{fig5A}, and  \ref{fig6A}, respectively. The  color-bar  indicates  the  total EM  included within a log T range shown in the upper left corner of the each panel. The DEM maps show that  campfires are visible at coronal temperatures (a,b,d,e,g,h,j,k). However, in panels (c, f, i, and l) there is barely any emission at temperatures above log$_{10}$ T(K) = 6.5. 
	} \label{fig8A}
\end{figure} 

\end{document}